\documentclass[11pt]{article}
\usepackage[letterpaper,margin=0.90in]{geometry}
\usepackage{amsmath,amsthm,amssymb}
\usepackage{graphicx}
\usepackage{pstricks,pstricks-add}
\usepackage{hyperref}
\usepackage{ifpdf}
\ifpdf
 \usepackage{pst-pdf}
\fi

\newtheorem{theorem}{Theorem}
\newtheorem{corollary}{Corollary}
\newtheorem{lemma}{Lemma}
\newtheorem{proposition}{Proposition}

\theoremstyle{definition}
\newtheorem{definition}{Definition}

\theoremstyle{remark}

\newcommand{\scsum}{\text{\sc Sum}}

\newcommand{\opt}{\text{\sc opt}}

\newcommand{\floor}[1]{\lfloor #1 \rfloor}
\newcommand{\half}{\frac{1}{2}}

\title{On the performance of approximate equilibria in congestion games}

\author{George Christodoulou\thanks{Max-Planck-Institut f\"{u}r Informatik, Saarbr\"{u}cken, Germany.
 Email: \texttt{\{gchristo\}@mpi-inf.mpg.de}} 
 \and 
 Elias Koutsoupias\thanks{Department of Informatics,
 University of Athens. Email: \texttt{elias@di.uoa.gr}}
\and 
 Paul G. Spirakis
 \thanks{Computer Engineering and Informatics Department, Patras University, Greece.
 Email: \texttt{\{spirakis\}@cti.gr}}
  }
\date{}

\begin{document}
\maketitle
\vspace*{-0.4in}
\begin{abstract}
We study the performance of approximate Nash equilibria for linear
congestion games. We consider how much the price of anarchy worsens
and how much the price of stability improves as a function of the
approximation factor $\epsilon$. We give (almost) tight upper and
lower bounds for both the price of anarchy and the price of stability
for atomic and non-atomic congestion games. Our results not only
encompass and generalize the existing results of exact equilibria to
$\epsilon$-Nash equilibria, but they also provide a unified approach
which reveals the common threads of the atomic and non-atomic price of
anarchy results. By expanding the spectrum, we also cast the existing
results in a new light. For example, the Pigou network, which gives
tight results for exact Nash equilibria of selfish routing, remains
tight for the price of stability of $\epsilon$-Nash equilibria.
\end{abstract}


\section{Introduction}

A central concept in Game Theory is the notion of equilibrium and in
particular the notion of Nash equilibrium.  Algorithmic Game Theory
has studied extensively and with remarkable success the computational
issues of Nash equilibria. As a result, we understand almost
completely the computational complexity of exact Nash equilibria (they
are PPAD-complete for games described explicitly \cite{DGP06,CD06} and
PLS-complete for games described succinctly \cite{FPT04}). The results
established a long suspected drawback of Nash equilibria, namely that
they cannot be computed effectively, thus upgrading the importance of
approximate Nash equilibria. We don't understand completely the
computational issues of approximate Nash equilibria
\cite{LMM03,FPT04,DMP06,TS07}, but they provide a more reasonable
equilibrium concept: It makes sense to assume that an agent is willing
to accept a situation that is almost optimal to him.

In another direction, a large body of research in Algorithmic Game
Theory concerns the degree of performance degradation of systems due
to the selfish behavior of its users. Central to this area is the
notion of price of anarchy (PoA) \cite{KP99,Pap01} and the price of
stability (PoS) \cite{ADKTWR04}. The first notion compares the social
cost of the worst-case equilibrium to the social optimum, which could
be obtained if every agent followed obediently a central
authority. The second notion is very similar but it considers the best
Nash equilibrium instead of the worst one.

A natural question then is how the performance of a system is affected
when its users are approximately selfish: What is the {\em approximate
  price of anarchy} and the {\em approximate price of stability}?
Clearly, by allowing the players to be almost rational (within an
$\epsilon$ factor), we expand the equilibrium concept and we expect
the price of anarchy to get worse. On the other hand, the price of
stability should improve. The question is how they change as functions
of the parameter $\epsilon$. This is exactly the question that we
address in this work.

We study two fundamental classes of games: the class of congestion
games \cite{Ros73,MS96} and the class of non-atomic congestion games
\cite{Mil96}.  The latter class of games includes the selfish routing
games which played a central role in the development of the price of
anarchy \cite{RT02,RT04}. The former class played also an important
role in the development of the area of the price of anarchy, since it
relates to the task allocation problem, which was the first problem to
be studied within the framework of the price of anarchy \cite{KP99}.
Although the price of anarchy and stability of these games for exact
equilibria was established long ago \cite{RT02,CK05a,CK05b,AAE05}---and
actually Tim Roughgarden \cite{Rou02,Rou05} addressed partially the
question for the price of anarchy of approximate equilibria---our
results add an unexpected understanding of the issues involved.

While these two classes of games are conceptually very similar,
dissimilar techniques were employed to answer the questions concerning
the PoA and PoS. Moreover, the qualitative aspects of the answers were
quite different. For instance, the Pigou network of two parallel links
captures the hardest network situation for the price of anarchy for
the selfish routing. In fact, Roughgarden \cite{Rou03}  proved that the
Pigou network is the worst case scenario for a very broad class of
delay functions. On the other hand, the lower bound for congestion
games is different and somewhat more involved \cite{AAE05,CK05a}.

For the selfish routing games, the price of stability is not different
than the price of anarchy because these games have a unique Nash (or
Wardrop as it is called in these games) equilibrium. On the other
hand, for the atomic congestion games, the problem proved more
challenging \cite{CK05b,CFKKM06}. New techniques exploiting the
potential of these games needed in order to come up with an upper
bound. The lower bound is quite complicated and, unlike the selfish
routing case, it has a dependency on the number of players (it attains
the maximum value at the limit).

The main difference between the two classes is the ``integrality'' of
atomic congestion games: In congestion games, when a player considers
switching to another strategy, he has to take into account the extra
cost that he will add to the edges (or facilities) of the new
strategy. The number of players on the new edges increases by one and
this changes the cost. On the other hand, in the selfish routing games
the change of strategies has no additional cost. A simple---although
not entirely rigorous---way to think about it, is to consider the
effects of a tiny amount of flow that ponders whether to change path:
it will not really affect the flow on the new edges (at least for
continuous cost functions).

Is integrality the reason which lies behind the difference of these
two classes of games? It seems so for the exact case. But our work
could be interpreted as revealing that the uniqueness of the Nash
equilibrium is also an important factor. Because when we move to the
wider class of $\epsilon$-Nash equilibria, the uniqueness is
dropped and the problems look quite similar qualitatively; the
integrality difference is still there, but it only manifests itself in
different quantitative or algebraic differences.

\subsection{Our contribution and related work}

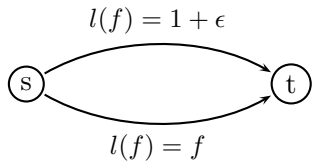
\begin{figure}
\centering
\begin{pspicture}(-2,-1)(2,1)
\psmatrix[mnode=circle,colsep=3cm]
 s & t
 \psset{shortput=nab,nodesep=1pt,arrows=->,labelsep=3pt}
 \small
 \ncarc[arcangle=30]{1,1}{1,2}^{$l(f)=1+\epsilon$}
 \ncarc[arcangle=-30]{1,1}{1,2}_{$l(f)=f$}
\endpsmatrix
\end{pspicture}
\caption{The Pigou network.}
\label{fig:pigou}
\end{figure}

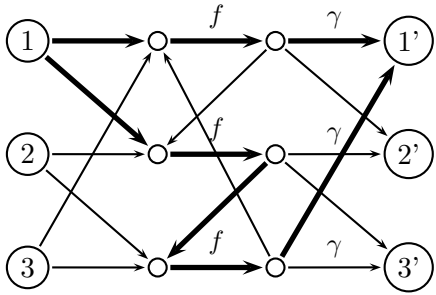
\begin{figure}
\centering
\begin{pspicture}(-3,-0.3)(3,4)
\psmatrix[mnode=circle,colsep=1.3cm,rowsep=0.8cm]
 1 & \ & \ &  1' \\
 2 & \ & \ &  2' \\
 3 & \ & \ &  3' 
 \psset{shortput=nab,nodesep=1pt,arrows=->,labelsep=3pt}
 \small
 \ncline[linewidth=2pt]{1,1}{1,2}
 \ncline[linewidth=2pt]{1,2}{1,3}^{$f$}
 \ncline[linewidth=2pt]{1,3}{1,4}^{$\gamma$}
 \ncline[linewidth=2pt]{1,1}{2,2}
 \ncline[linewidth=2pt]{2,3}{3,2}
 \ncline[linewidth=2pt]{3,3}{1,4}

 \ncline{2,1}{2,2}
 \ncline[linewidth=2pt]{2,2}{2,3}^{$f$}
 \ncline{2,3}{2,4}^{$\gamma$}
 \ncline{2,1}{3,2}
 \ncline{3,3}{1,2}
 \ncline{1,3}{2,4}

 \ncline{3,1}{3,2}
 \ncline[linewidth=2pt]{3,2}{3,3}^{$f$}
 \ncline{3,3}{3,4}^{$\gamma$}
 \ncline{3,1}{1,2}
 \ncline{1,3}{2,2}
 \ncline{2,3}{3,4}
\endpsmatrix
\end{pspicture}
\caption{Lower bound for selfish routing. There are 3 distinct edge
  latency functions: $l(f)=f$, $l(f)=\gamma$ (a constant which depends
  on $\epsilon$), $l(f)=0$ (omitted in the picture). There are 3
  commodities of rate 1 with source $i$ and destination $i'$. The two
  paths for the first commodity are shown in bold lines.}
\label{fig:routing_lb}
\end{figure}

Our work encompasses and generalizes some fundamental results in the
area of the price of anarchy \cite{RT02,CK05a,CK05b,AAE05} (see also
the recently published book \cite{NRTV07} for background
information). Our techniques not only provide a unifying approach but
they cast the existing results in a new light. For instance, the Pigou
network (Figure~\ref{fig:pigou}) is still the tight example for the
price of stability, but not the price of anarchy. Instead for the
price of anarchy, the network of Figure~\ref{fig:routing_lb} is tight;
in fact, this network is tight only for $\epsilon\leq 1$; a more
complicated network is required for larger $\epsilon$.

We consider $\epsilon$ approximate Nash equilibria. We use the
\emph{multiplicative definition of approximate equilibria}: In
congestion games, a player does not switch to a new strategy as long
as his current cost is less than $1+\epsilon$ times the new cost. In
the selfish routing games, we use exactly the same definition: the
flow is at an equilibrium when the cost on its paths is less than
$1+\epsilon$ times the cost of every alternate path.  There have been
other definitions for approximate Nash equilibria in the
literature. The most-studied is the additive case \cite{LMM03,
  DGP06}. In \cite{CS07}, they consider approximate equilibria of the
multiplicative case and they study convergence issues for congestion
games. Our definition differs slightly and our results can be
naturally adapted to the definition of \cite{CS07}.

There is a large body of work on the price of anarchy in various
models \cite{NRTV07}. More relevant to our work are the following
publications: In \cite{AAE05,CK05a}, it is proved that the price of
anarchy of congestion games for pure equilibria is
$\frac{5}{2}$. Later in \cite{CK05b}, it is showed that the ratio
$5/2$ is tight even for correlated equilibria, and consequently for
mixed equilibria. For symmetric games, it is something less:
$\frac{5n-2}{2n+1}$ \cite{CK05a}, where $n$ is the number of
players. For weighted congestion games, the price of anarchy is
$1+\phi\approx 2.618$ \cite{AAE05}. Later in \cite{CK05b}, it was
proved that the same ratio holds even for correlated equilibria. In
\cite{CK05b, CFKKM06}, it was shown that the price of stability of
linear congestion games is $1+\sqrt{3}/3$. For the selfish routing
paradigm, the price of anarchy (and of stability) for linear latencies
is $4/3$ \cite{RT02} (see also \cite{CSM04a} for a simplified version
of this proof), and the results are extended to non-atomic games in
\cite{RT04}. The most relevant work is \cite{Rou02,Rou05} which gives
tight bounds for the price of anarchy of approximate equilibria when
$\epsilon\leq 1$. We extend this to every positive $\epsilon$ using
different techniques.

In this work, we give (almost) tight upper and lower bounds for the
PoA and PoS of atomic and non-atomic linear congestion games. Our
results are summarized in the Table~\ref{tab:results} (where atomic
refers to congestions games).

\begin{table}[htbp]
\centering
\begin{tabular}{|c||c|c|}
\hline
           & Atomic & Non-atomic \\
\hline\hline
Anarchy &
\begin{minipage}{0.3\linewidth}
$(1+\epsilon)\frac{z^2+3z+1}{2z-\epsilon}$ \\ where
$z=\lfloor\frac{1+\epsilon+\sqrt{5+6\epsilon+\epsilon^2}}{2}\rfloor$ \\
(Section 3)
\end{minipage}
& \begin{minipage}{0.3\linewidth}
$(1+\epsilon)^2$ \\
(especially for $\epsilon\leq 1$: $\frac{4(1+\epsilon)}{3-\epsilon}$) \\
(Section 4)
\end{minipage} \\
\hline
Stability  & \begin{minipage}{0.3\linewidth}$\frac{1+\sqrt{3}}{\epsilon+\sqrt{3}}$ \\
(Section 5) \end{minipage} & \begin{minipage}{0.3\linewidth}$\frac{4}{(3-\epsilon)(1+\epsilon)}$ \\
(Section 6) \end{minipage} \\
\hline
\end{tabular}
\caption{The upper bounds (with pointers to relevant sections).} 
\label{tab:results}
\end{table}

The results in the above table include the upper bounds. We have
matching lower bounds except for the atomic PoS (and for non-integral
values $\epsilon>1$ for the PoA of non-atomic case). The price of
stability reduces to 1 for $\epsilon\geq 1$, which means that the
optimal is a $1$-Nash equilibrium, for both the atomic and non-atomic
case. Also, the price of anarchy is approximately
$(1+\epsilon)(3+\epsilon)$ and $(1+\epsilon)^2$ for large $\epsilon$,
the atomic and non-atomic case respectively. The price of anarchy for
$\epsilon\leq 1$ has been established before in \cite{Rou02,Rou05}.

The interesting case is probably when $\epsilon$ is small. For
$\epsilon\leq 1/3$ the results are summarized in the
Table~\ref{tab:results2}:

\begin{table}[htbp]
\centering
\begin{tabular}{|c||c|c|}
\hline
           & Atomic & Non-atomic \\
\hline\hline
Anarchy &
$\frac{5(1+\epsilon)}{2-\epsilon}$
& $\frac{4(1+\epsilon)}{3-\epsilon}$ \\
\hline
Stability  & $\frac{1+\sqrt{3}}{\epsilon+\sqrt{3}}$ & $\frac{4}{(3-\epsilon)(1+\epsilon)}$ \\
  \hline
\end{tabular}
\caption{The upper bounds for $\epsilon\leq 1/3$.} 
\label{tab:results2}
\end{table}

A useful tool, interesting in its own right, is a generalization of
the notion of potential function for both the atomic and non-atomic
case (Theorems \ref{thm:potential-atomic} and
\ref{thm:potential-selfish}) for the case of $\epsilon$-Nash
equilibria. Remarkably, the parameter $\epsilon$ appears only in the
linear part of the (quadratic) potential function.

Our approach is similar to \cite{CK05a,CK05b}, but it is much more
involved technically and requires a deeper understanding of the
potential function issues involved. We want also to draw attention to
our techniques in bounding the approximate price of anarchy for the
selfish routing which differ considerably from the techniques of
\cite{RT02} and others \cite{NRTV07}. The main difference is that we
move from a domain with unique equilibrium to a domain with a set of
solutions.


\section{Definitions}


A congestion game~\cite{Ros73}, also called an exact potential game
\cite{MS96}, is a tuple $(N,E,(\mathcal{S}_i)_{i\in N},(f_e)_{e\in
  E})$, where $N=\{1,\ldots,n\}$ is a set of $n$ players, $E$ is a set
of facilities, $\mathcal{S}_i\subseteq 2^E$ is a set of pure
strategies for player $i$: a pure strategy $A_i\in \mathcal{S}_i$ is
simply a subset of facilities and $l_e$ is a cost (or latency)
function, one for each facility $e\in E$. The cost of player $i$ for
the pure strategy profile $A=(A_1,\ldots, A_n)$ is $c_i(A)=\sum_{e\in
  A_i}l_e(n_e(A))$, where $n_e(A)$ is the number of players who use
facility $e$ in the strategy profile $A$. 

\begin{definition}
A pure strategy profile $A$ is an $\epsilon$ equilibrium iff for every
player $i\in N$
\begin{equation}\label{eq:approximate_inequality}
c_i(A)\leq (1+\epsilon)c_i(A_i,A_{-i}), \qquad \forall A_i\in \mathcal{S}_i
\end{equation}
\end{definition}
We believe that the multiplicative definition of approximate
equilibria makes more sense in the framework that we consider. This is
because the costs of the players usually vary in this setting and a
uniform $\epsilon$ does not make much sense. Given that the price of
anarchy is a ratio, we need a definition that is insensitive to
scaling.

The social cost of a pure strategy profile $A$ is the sum of the
players cost
$$SC(A)=\scsum(A)=\sum_{i\in N}c_i(A)$$

The pure approximate price of anarchy, is the social cost of the worst
case $\epsilon$ equilibrium over the optimal social cost

$$PoA = \max_{A \text{ is a $\epsilon$-Nash}}\frac{SC(A)}{\opt},$$

while the pure approximate price of stability, is the social cost of
the worst case $\epsilon$ equilibrium over the optimal social cost

$$PoS = \min_{A \text{ is a $\epsilon$-Nash}}\frac{SC(A)}{\opt}.$$

Instead of defining formally the class of nonatomic congestion games,
we prefer to focus on the more illustrative--more restrictive
though--class of selfish routing games. The difference in the two
models is that in a non-atomic game, there does not exist any network
and the strategies of the players are just subsets of facilities (as
in the case of atomic congestion games) and they do not necessarily
form a path in a network. The most desirable results are obtained when
the upper bounds hold for general non-atomic games and matching lower
bounds hold for the special case of selfish routing. Our results
almost follow this pattern, with a few exceptions of lower
bounds. This is because we put emphasis on the simplicity and we
didn't attempt to extend them to the selfish routing case.

Let $G=(V,E)$ be a directed graph, where $V$ is a set of vertices and
$E$ is a set of edges. In this network we consider $k$ commodities:
source-node pairs $(s_i, t_i)$ with $i=1\ldots k$, that define the
sources and destinations. The set of simple paths in every pair
$(s_i,t_i)$ is denoted by $\mathcal{P}_i$, while with
$\mathcal{P}=\cup_{i=1}^k\mathcal{P}_i$ we denote their union. A flow
$f$, is a mapping from the set of paths to the set of nonnegative
reals $f:\mathcal{P}\rightarrow\mathbb{R}^{+}$. For a given flow $f$,
the flow on an edge is defined as the sum of the flows of all the
paths that use this edge $f_e=\sum_{P\in \mathcal{P}, e\in P}f_P$. We
relate with every commodity $(s_i, t_i)$ a traffic rate $r_i$, as the
total traffic that needs to move from $s_i$ to $t_i$. A flow $f$ is
feasible, if for every commodity $\{s_i, t_i\}$, the traffic rate
equals the flow of every path in $\mathcal{P}_i$, $r_i=\sum_{P\in
  \mathcal{P}_i}{f_P}$. Every edge introduces a delay in the
network. This delay depends on the load of the edge and is determined
by a delay function, $l_e(\cdot)$. An instance of a routing game is
denoted by the triple $(G,r,l)$. The latency of a path $P$, for a
given flow $f$, is defined as the sum of all the latencies of the
edges that belong to $P$, $l_P(f)=\sum_{e\in P}l_e(f_e)$. The social
cost that evaluates a given flow $f$, is the total delay due to $f$
$$C(f)=\sum_{P\in \mathcal{P}}l_P(f)f_P.$$

The total delay can also be expressed via edge flows 
$C(f)=\sum_{e\in E}l_e(f_e)f_e.$

From now on, when we are talking about flows, we mean feasible flows.
In \cite{BMW56,DS69}, it is shown that there exists a (unique)
equilibrium flow, known as Wardrop equilibrium\cite{War52}. In analogy
to their definition, we define the $\epsilon$ Wardrop equilibrium
flows, as follows

\begin{definition}\label{def:selfish-approx}
A feasible flow $f$, is an $\epsilon$-Nash (or Wardrop) equilibrium,
if and only if for every commodity $i\in \{1,\ldots ,k\}$ and
$P_1,P_2\in \mathcal{P}_i$ with $f_{P_1}>0$, $l_{P_1}(f)\leq
(1+\epsilon)l_{P_2}(f)$.
\end{definition}

In this work we restrict our attention to \emph{linear latency
  functions}: $l_e(x)=a_e x+b_e$, where $a_e$ and $b_e$ are
nonnegative constants. Our results naturally extend to mixed and
correlated equilibria. We also believe that they can be also extended
to more general latency functions such as polynomials.


\section{Congestion Games --  PoA}

In this section we study the dependency on the parameter $\epsilon$,
of the price of anarchy for the case of atomic congestion games.  For
large $\epsilon$ the price of anarchy is roughly $(1+\epsilon)^2$. The
same holds the non-atomic case as we are going to establish in the
next section.

We will need the following arithmetic lemma. 

\begin{lemma}\label{lem:poa-upper-atomic}
For every $\alpha,\beta,z\in \mathbb{N}$:
$$\beta (\alpha + 1)\leq \frac{1}{2z+1}\alpha^2  + 
\frac{z^2+3z+1}{2z+1}\beta^2$$
\end{lemma}

\begin{proof}
Consider the function $f(\alpha,\beta)$ which we obtain when we
subtract the left part of the statement's inequality from the right
part and multiply the result by $2z+1$.

\begin{eqnarray*}
f(\alpha,\beta) &= & a^2+(z^2+3z+1)\beta^2-(2z+1)\beta (\alpha+1)\\
&=& \left(\alpha-\frac{2z+1}{2}\beta\right)^2+\frac{(8z+3)\beta^2-(8z+4)\beta}{4}.
\end{eqnarray*}
For $\beta=0$, and for any $\beta \geq 2$, $f(\alpha,\beta)$ is
clearly positive. For $\beta=1$ it takes the form
$f(\alpha,1)=(\alpha-z)(\alpha-z-1)\geq 0$, and the lemma follows.
\end{proof}

Our first theorem gives an upper bound for the price of anarchy for
congestion games; this is tight, as we are going soon to
establish. This result generalizes the bound in \cite{AAE05,CK05a} to
approximate equilibria. The proof is for linear latency functions of
the form $l_e(x)=x$, but it can be easily extended to latencies of the
form $l_e(x)=a_ex+b_e$, with nonnegative $a_e,b_e$.

\begin{theorem}[Atomic-PoA-Upper-Bound]\label{thm:poa-upper-atomic}
For any positive real $\epsilon$, the approximate price of anarchy of
general congestion games with linear latencies is at most
$$(1+\epsilon)\frac{z^2+3z+1}{2z-\epsilon},$$
where $z\in \mathbb{N}$ is the maximum integer with
$\frac{z^2}{z+1}\leq 1+\epsilon$ (or equivalently for
$z=\lfloor\frac{1+\epsilon+\sqrt{5+6\epsilon+\epsilon^2}}{2}\rfloor$).
\end{theorem}
\begin{proof}
Let $A=(A_1,\ldots, A_n)$ be an $\epsilon$-approximate pure Nash, and
$P=(P_1,\ldots, P_n)$ be the optimum allocation.  From the definition
of $\epsilon$-equilibria (Inequality
(\ref{eq:approximate_inequality})) we get

$$\sum_{e\in A_i}n_e(A)\leq \left(1+\epsilon\right)\sum_{e\in P_i}\left(n_e(A)+1\right).$$

If we sum up for every player $i$ and use
Lemma~\ref{lem:poa-upper-atomic}, we get

\begin{eqnarray*}
\scsum(A) &=& \sum_{i \in N}c_i(A)\\
		&=& \sum_{i \in N}\sum_{e\in A_i}n_e(A)\\
		&=& \sum_{e\in E}n_e^2(A)\leq \left(1+\epsilon\right)\sum_{e\in E}n_e(P)\left(n_e(A)+1\right)\\
		&\leq & \frac{1+\epsilon}{2z+1}\sum_{e\in E}n_e^2(A)+\frac{(1+\epsilon)(z^2+3z+1)}{2z+1}\sum_{e\in E}n_e^2(P)\\
		&=& \frac{1+\epsilon}{2z+1}\scsum(A)+\frac{(1+\epsilon)(z^2+3z+1)}{2z+1}\opt.
\end{eqnarray*}
From this we obtain the theorem
$$\scsum(A)\leq (1+\epsilon)\frac{z^2+3z+1}{2z-\epsilon}\opt.$$
\end{proof}

The above is a typical proof in this work. All our upper bound proofs
have similar form. The proofs of the price of stability are more
challenging however, as they require the use of appropriate
generalizations of the potential function. We now show that the above
upper bound is tight.

\begin{theorem}[Atomic-PoA-Lower-Bound]\label{thm:poa-lower-atomic}
For any real positive $\epsilon$, there are instances of congestion
games with linear latencies, for which the approximate price of
anarchy of general congestion games with linear latencies, is at
least $$(1+\epsilon)\frac{z^2+3z+1}{2z-\epsilon},$$ where $z\in
\mathbb{N}$ is the maximum integer with $\frac{z^2}{z+1}\leq
1+\epsilon$.
\end{theorem}

\begin{proof}
Let $z\in \mathbb{N}$ be the maximum integer with $\frac{z^2}{z+1}\leq
1+\epsilon$. We will construct an instance with $z+2$ players and
$2z+4$ facilities. There are two types of facilities:
\begin{itemize}
\item $z+2$ facilities of type $\alpha$, with latency $l_e(x)=x$ and 
\item $z+2$ facilities of type $\beta$ with latency $l_e(x)=\gamma
x=\frac{(z+1)^2-(1+\epsilon)(z+2)}{(1+\epsilon)(z+1)-z^2}x$.
\end{itemize}
Player $i$ has two alternative pure strategies, $S_i^1$ and $S_i^2$.
\begin{itemize}
\item The first strategy is to play the two facilities $\alpha_i$ and
$\beta_i$, i.e. $S_i^1=\{\alpha_i,\beta_i\}$.
\item The second strategy is to play every facility of type $\alpha$
except for $\alpha_i$ and $z+1$ facilities of type $\beta$ starting at
facility $\beta_{i+1}$. More precisely, the second strategy has the
facilities  
$$S_i^2=\{\alpha_1,\ldots, \alpha_{i-1},\alpha_{i+1},\ldots
,\alpha_{z+2},\beta_{i+1},\ldots ,\beta_{i+1+z}\},$$ where the indices
may require computations $(\mod z+2)$.
\end{itemize}

First we prove that playing the second strategy $S^2=(S_1^2,\ldots
,S_n^2)$ is a $\epsilon$-Nash equilibrium. The cost of player $i$
is $$c_i(S^2)=(z+1)^2+\gamma z^2,$$ as there are exactly $z+1$ players
using facilities of type $\alpha$ and exactly $z$ players using
facilities of type $\beta$.

If player $i$ unilaterally switches to the other available strategy
$S^1_i$ he has cost
$$c_i(S_i^1,S^2_{-i})=(z+2)+\gamma(z+1)=\frac{c_i(S^2)}{1+\epsilon},$$
which shows that $S^2$ is an $\epsilon$-Nash equilibrium.

The optimum allocation is the strategy profile $S^1$, where every
player has cost $c_i(S^1)=1+\gamma$ and so the price of anarchy is
$$\frac{c_i(S^2)}{c_i(S^1)}=\frac{(z+1)^2+\gamma z^2}{1+\gamma}= 
(1+\epsilon)\frac{z^2+3z+1}{2z-\epsilon}.$$ Notice that the parameter
$z$ is an integer because it expresses a number of facilities.
\end{proof}

The above theorems (lower and upper bound) employ, for any positive
real $\epsilon$, an integer $z(\epsilon)$, which is the maximum
integer that satisfies $\frac{z^2}{z+1}\leq 1+\epsilon$. So for
$\epsilon \in [0,1/3]$, $z(\epsilon)=1$ and the price of anarchy is
$\frac{5(1+\epsilon)}{2-\epsilon}$, for $\epsilon \in [1/3,5/4]$,
$z(\epsilon)=2$ and the price of anarchy is
$\frac{11(1+\epsilon)}{4-\epsilon}$ and so on. Roughly the price of
anarchy grows as $(1+\epsilon)(3+\epsilon)$.

\section{Selfish Routing -- PoA}
\label{sec:poa-selfish}

In this section we estimate the price of anarchy for non-atomic
congestion games and consequently for its special case, the selfish
routing. Our results generalize the results in \cite{RT02,RT04} to the
case of approximate equilibria. The proof has the same form with the
proof of the atomic case in the previous section.

Again, we will need an arithmetic lemma. The main change now is that
we deal with continuous values instead of integrals.

\begin{lemma}\label{lem:poa-upper-selfish}
For every reals $\alpha,\beta, \lambda$ it holds,
$$\beta \alpha \leq \frac{1}{4\lambda}\alpha^2  + \lambda\beta^2,$$ where 
\end{lemma}

\begin{proof}
Simply because
$\alpha^2+4\lambda^2\beta^2-4\lambda\alpha\beta=\left(\alpha-2\lambda\beta\right)^2\geq
0$.
\end{proof}

\begin{theorem}[Selfish-PoA-Upper-Bound]\label{thm:poa-upper-selfish}
For any positive real $\epsilon$, and for every $\lambda\geq 1$, the
approximate price of anarchy of non-atomic congestion games with
linear latencies is at most
$$\frac{4\lambda^2(1+\epsilon)}{4\lambda-1-\epsilon}.$$
\end{theorem}

\begin{proof}
Let $f$ be an $\epsilon$-approximate Nash flow, and $f^*$ be the
optimum flow (or any other feasible flow).  From the definition of
approximate Nash equilibria (Inequality
(\ref{eq:approximate_inequality})), we get that for every path $p$
with non-zero flow in $f$ and any other path $p'$:
$$\sum_{e\in p} l_e(f_e) \leq (1+\epsilon) \sum_{e\in p'}
l_e(f_e^*).$$ We sum these inequalities for all pairs of paths $p$ and
$p'$ weighted with the amount of flow of $f$ and $f^*$ on these paths.
\begin{align*}
\sum_{p,p'}f_p f_{p'}^* \sum_{e\in p} l_e(f_e) & \leq (1+\epsilon)
\sum_{p,p'}f_p f_{p'}^* \sum_{e\in p'}
l_e(f_e^*) \\
\sum_{p'} f_{p'}^* \sum_{e\in E} l_e(f_e)f_e & \leq (1+\epsilon)
\sum_{p} f_p \sum_{e\in E} l_e(f_e^*)f_e^* \\
(\sum_{p'} f_{p'}^*) \sum_{e\in E} l_e(f_e)f_e & \leq (1+\epsilon)
(\sum_{p} f_p) \sum_{e\in E} l_e(f_e^*)f_e^*
\end{align*}
But $\sum_p f_p=\sum_{p'} f_{p'}^*$ is equal to the total rate for the
feasible flows $f$ and $f^*$. Simplifying, we get 
$$\sum_{e\in E}l_e(f_e)f_e \leq (1+\epsilon)\sum_{e\in E}l_e(f_e)f_e^*.$$
This is the generalization to approximate equilibria of the inequality
established by Beckmann, McGuire, and Winston \cite{BMW56} for exact
Wardrop equilibria.

Since we consider linear functions of the form $l_e(f_e)=a_ef_e+b_e$,
we get
$$\sum_{e\in E}\left(a_ef_e^2+b_ef_e\right) \leq (1+\epsilon)\sum_{e\in E}a_ef_ef_e^* + (1+\epsilon)\sum_{e\in E}b_ef_e^*.$$

Applying Lemma \ref{lem:poa-upper-selfish}, we get
$$\sum_{e\in E}\left(a_ef_e^2+b_ef_e\right) \leq (1+\epsilon)\sum_{e\in E}a_e\left(\frac{1}{4\lambda}f_e^2+\lambda{f^*_e}^2\right) + (1+\epsilon)\sum_{e\in E}b_ef_e^*.$$
from which we get
$$\sum_{e\in E}\left(a_e\left(1-(1+\epsilon)\frac{1}{4\lambda}\right)f_e^2+b_ef_e\right) \leq \lambda(1+\epsilon)\sum_{e\in E}a_e {f^*_e}^2 + (1+\epsilon)\sum_{e\in E}b_ef_e^*,$$ 
and for $\lambda \geq 1$
$$\frac{4\lambda-1-\epsilon}{4\lambda}SC(f)\leq (1+\epsilon)\lambda SC(f^*).$$

This gives price of anarchy at most
of $$\frac{4\lambda^2(1+\epsilon)}{4\lambda-1-\epsilon},$$
for every $\lambda\geq 1$.
\end{proof}

The expression $\frac{4\lambda^2(1+\epsilon)}{4\lambda-1-\epsilon}$ of
the theorem is minimized for $\lambda=(1+\epsilon)/2$ when
$\epsilon\geq 1$ (and $\lambda=1$ when $\epsilon\leq 1$). We therefore
obtain the following two corollaries by substituting $\lambda=1$ and
$\lambda=(1+\epsilon)/2$. The first corollary was proved before in
\cite{Rou02,Rou05} using different techniques.

\begin{corollary}\label{cor:poa-upper-selfish}
For any nonnegative real $\epsilon\leq 1$, the approximate price of
anarchy of non-atomic congestion games with linear latencies is at
most
$$\frac{4(1+\epsilon)}{3-\epsilon}.$$
\end{corollary}

\begin{corollary}\label{cor:poa-upper-selfish-large}
For any positive real $\epsilon\geq 1$, the approximate price of
anarchy of non-atomic congestion games with linear latencies is at
most
$$\left(1+\epsilon\right)^2.$$ 
\end{corollary}

We now show that the above upper bounds are tight. To be precise, we
show that Corollary~\ref{cor:poa-upper-selfish} is tight and that
Corollary~\ref{cor:poa-upper-selfish-large} is partially tight---only
for integral values of $\epsilon$.

The following theorem for the case of $\epsilon\leq 1$ was first shown
in \cite{Rou02,Rou05}. We include a different proof here for
completeness and because it is similar to the generalization for
$\epsilon>1$, in Theorem \ref{thm:poa-lower-selfish-large}.

\begin{theorem}[Selfish-PoA-Lower-Bound for $\epsilon\leq
1$]\label{thm:poa-lower-selfish}
For any nonnegative real $\epsilon\leq 1$, there are instances of
congestion games with linear latencies, for which the approximate
price of anarchy of general congestion games with linear latencies, is
at least $$\frac{4(1+\epsilon)}{3-\epsilon}.$$
\end{theorem}

\begin{proof}
We will construct an instance with $3$ commodities, each of them with
unit flow, and $6$ facilities (a slightly more involved network case
appears in Figure~\ref{fig:routing_lb}. There are two types of
facilities:
\begin{itemize}
\item $3$ facilities of type $\alpha$, with latency $l(x)=x$ and 
\item $3$ facilities of type $\beta$ with constant latency $l(x)=\gamma=\frac{2(1-\epsilon)}{1+\epsilon}$.
\end{itemize}
Commodity $i$ has two alternative pure strategies, $S_i^1$ and
$S_i^2$.
\begin{itemize}
\item The first strategy is to choose both the facilities $\alpha_i$
and $\beta_i$, i.e. $S_i^1=\{\alpha_i,\beta_i\}$
\item As a second alternative, players of commodity $i$ may choose
every facility of type $\alpha$ except for $\alpha_i$; we denote this
set by $S_i^2=\{\alpha_{-i}\}$.
\end{itemize}
First we prove that playing the second strategy
$S^2=(S_1^2,S_2^2,S_3^2)$ is a $\epsilon$-Nash equilibrium. The cost
of every player in commodity $i$ is $c_i(S^2)=4$, as there are exactly
$z+1$ players using facilities of type $\alpha$ and exactly $z$
players using facilities of type $\beta$.

If player $i$ unilaterally switches to the other available strategy
$S^1_i$ he gets
$$c_i(S_i^1,S^2_{-i})=2+\gamma=\frac{c_i(S^2)}{1+\epsilon}$$ and so 
$S^2$ is an $\epsilon-$approximate equilibrium.

In the optimum case, the players use strategy profile $S^1$, where
commodity $i$ has cost $c_i(S^1)=1+\gamma$ and so the price of anarchy
is
$$\frac{c_i(S^2)}{c_i(S^1)}=\frac{4}{1+\gamma}=\frac{4(1+\epsilon)}{3-\epsilon}.$$
\end{proof}

For larger $\epsilon$ ($\epsilon>1$), we have:
\begin{theorem}[Selfish-PoA-Lower-Bound for $\epsilon\geq
1$]\label{thm:poa-lower-selfish-large}
For any real positive $\epsilon$, there are instances of congestion
games with linear latencies, for which the approximate price of
anarchy of general congestion games with linear latencies, is at
least $$(1+\epsilon)\frac{z(z+1)}{2z-\epsilon}=(1+\epsilon)\frac{z^2+z}{2z-\epsilon},$$
where $z =\floor{1+\epsilon}.$
\end{theorem}

\begin{proof}
Let $z=\floor{1+\epsilon}$.  We will construct an instance with $z+2$
commodities and $2z+4$ facilities. There are two types of facilities:
\begin{itemize}
\item $z+2$ facilities of type $\alpha$, with latency $1$ and
\item $z+2$ facilities of type $\beta$ with latency
$\gamma=\frac{(z+1)^2-(1+\epsilon)(z+1)}{(1+\epsilon)z-z^2}$.
\end{itemize}
Commodity $i$ has two alternative pure strategies, $S_i^1$ and
$S_i^2$.
\begin{itemize}
\item The first strategy is to choose both the facilities $\alpha_i$
and $\beta_i$, i.e. $S_i^1=\{\alpha_i,\beta_i\}$.
\item As a second alternative, commodity $i$ may choose every facility
of type $\alpha$ except for $\alpha_i$ and $z$ facilities of type
$\beta$ as defined in the following:
$$S_i^2=\{\alpha_1,\ldots, \alpha_{i-1},\alpha_{i+1},\ldots
,\alpha_{z+2},\beta_{i+1},\ldots ,\beta_{z+i+1}\},$$ where the indices
are computed $(\mod z+2)$.
\end{itemize}
First we prove that playing the second strategy $S^2=(S_1^2,\ldots
,S_n^2)$ is a $\epsilon$-Nash equilibrium.  The cost of commodity $i$
is $$c_i(S^2)=(z+1)^2+\gamma z^2,$$ as there are exactly $z+1$
commodities using facilities of type $\alpha$ and exactly $z$ players
using facilities of type $\beta$.

If commodity $i$ unilaterally switches to the other available strategy
$S^1_i$, its cost becomes
$$c_i(S_i^1,S^2_{-i})=(z+1)+\gamma z=\frac{c_i(S^2)}{1+\epsilon},$$
which shows that $S^2$ is an $\epsilon-$approximate equilibrium.

The optimum is the strategy profile $S^1$, where every commodity has
cost $c_i(S^1)=1+\gamma$. It follows that the price of anarchy is
$$\frac{c_i(S^2)}{c_i(S^1)}=\frac{(z+1)^2+\gamma z^2}{1+\gamma}=(1+\epsilon)\frac{z(z+1)}{2z-\epsilon}.$$
\end{proof}


\section{Atomic Games -- PoS}

An upper bound of the price of stability is perhaps more difficult to
obtain because we have to find a way to characterize the best
$\epsilon$-Nash equilibrium. We don't know how to do this, so we use
an indirect approach: We identify a property such that every profile
satisfying the property is guaranteed to be an $\epsilon$-Nash equilibrium. We then
upper bound the price of anarchy of all profiles satisfying this
property. To this end, we generalize the notion of potential
\cite{MS96}; a characteristic property of congestion games is that
they possess a potential function.

We define the $\epsilon$-potential function of a profile $A$ to be
q$$\Phi^\epsilon(A)=\half\sum_{e\in E} (a_e n_e(A)+b_e)
n_e(A)+\half\,\frac{1-\epsilon}{1+\epsilon}\sum_{e\in
  E}(a_e+b_e)n_e(A).$$ For $\epsilon=0$, this reduces to the classical
potential function for congestion games. More importantly, it
generalizes the following interesting property to $\epsilon$-Nash
equilibrium.
\begin{theorem}\label{thm:potential-atomic}
Any profile $A$ which is a local minimum of $\Phi^{\epsilon}$, is an
$\epsilon$-Nash equilibrium.
\end{theorem}
\begin{proof}
Consider a profile $A=(A_1,\ldots,A_n)$. We want to compute the change
in the $\epsilon$-potential function when player $i$ changes from
strategy $A_i$ to a strategy $P_i\in \mathcal{S}_i$. The resulting
profile $(P_i,A_{-i})$ has
$$n_e(P_i,A_{-i})=\left\{
\begin{array}{lc}
n_e(A)+1, & e\in P_i-A_i\\
n_e(A)-1, & e\in A_i-P_i\\
n_e(A), & \text{ otherwise. }
\end{array}
\right.
$$
From this we can compute the difference
\begin{align*}
\Phi^{\epsilon}(P_i,A_{-i})-\Phi^{\epsilon}(A)=&\sum_{e\in P_i-A_i}
\left(a_e n_e(A)+\frac{1}{1+\epsilon}(a_e+b_e)\right) - \\
& \sum_{e\in A_i-P_i} \left(a_e n_e(A)+\frac{1}{1+\epsilon}(-a_e
\epsilon +b_e)\right).
\end{align*}
We can rewrite this as
\begin{align}\label{eq:deviation}
\Phi^{\epsilon}(P_i,A_{-i})-\Phi^{\epsilon}(A)=&\sum_{e\in P_i}
\left(a_e n_e(A)+\frac{1}{1+\epsilon}(a_e+b_e)\right) - \sum_{e\in
  P_i\cap A_i} a_e - \\\nonumber
& \sum_{e\in A_i} \left(a_e n_e(A)+\frac{1}{1+\epsilon}(-a_e
\epsilon +b_e)\right).
\end{align}

Suppose now that profile $A$ is a local minimum of
$\Phi^{\epsilon}$. This translates to
$\Phi^{\epsilon}(P_i,A_{-i})\geq\Phi^{\epsilon}(A)$ for all $i$. The
cost for player $i$ before the change is $c_i(A)=\sum_{e\in A_i}
\left(a_e n_e(A)+b_e\right)$ and after the change is
$c_i(P_i,A_{-i})=\sum_{e\in P_i} \left(a_e
n_e(P_i,A_{-i})+b_e\right)$. We want to show that $A$ is an
$\epsilon$-Nash equilibrium: $c_i(A)\leq (1+\epsilon)c_i(P_i,A_{-i})$.

The $\epsilon$-potential consists of two parts that can be used to
bound the cost of player $i$ at profile $A$ and $(P_i,A_{-i})$:
\begin{align*}
c_i(A)&=\sum_{e\in A_i} (a_e n_e(A) + b_e) \\
& \leq \sum_{e\in A_i} (1+\epsilon)\left(a_e
n_e(A)+\frac{1}{1+\epsilon}(-a_e \epsilon +b_e)\right),
\end{align*}
(which holds because $n_e(A)\geq 1$ when $e\in A_i$), and
\begin{align*}
c_i(P_i,A_{-i})&=\sum_{e\in P_i}\left(a_e
(n_e(A)+1)+b_e\right)-\sum_{e\in
  P_i\cap A_i} a_e \\
&\geq \sum_{e\in P_i}\left(a_e n_e(A) +
\frac{1}{1+\epsilon}(a_e+b_e)\right)-\sum_{e\in P_i\cap A_i} a_e
\end{align*}
(which holds for $\epsilon\geq 0$).

It follows immediately that $c_i(A)\leq
(1+\epsilon)c_i(P_i,A_{-i})$. Consequently, $A$ is an $\epsilon$-Nash
equilibrium.
\end{proof}

First we present an easy upper bound, that uses only the previous
theorem.

\begin{proposition}
For linear congestion games, the price of stability is at most
$\frac{2}{1+\epsilon}$.
\end{proposition}

\begin{proof}
Let $A$ be the allocation that minimizes the $\epsilon$ potential
$\Phi^{\epsilon}$, and let $P$ be the optimum allocation.

We have $$\Phi^\epsilon(A)\leq \Phi^\epsilon(P)$$ and so
\begin{equation}\label{eq:eps-potential}
\scsum(A)+\frac{1-\epsilon}{1+\epsilon}\sum_{e\in E}(a_e+b_e)n_e(A)\leq \scsum(P)+\frac{1-\epsilon}{1+\epsilon}\sum_{e\in E}(a_e+b_e)n_e(P),
\end{equation}
from which we get 
$$\scsum(A)\leq \frac{2}{1+\epsilon}\scsum(P).$$
\end{proof}

The previous theorem gives us an easy way to bound the price of
stability. Clearly this is not tight: for $\epsilon=0$, it doesn't
provide us the right answer $1+\sqrt{3}/3$~\cite{CK05b,CFKKM06},
although it gives us a good estimation. To get a better upper bound we
need to work harder.

\begin{lemma}\label{lem:eps-pos}
For integers $\alpha,\beta$ and for $\gamma={\frac { \left( 3+2\,\sqrt
    {3} \right) \left( e-3+2\,\sqrt {3} \right) }{3\,e+3+2\,\sqrt
    {3}}}$
$$\gamma\beta^2 + \frac{1-\gamma\epsilon}{1+\epsilon}\beta-\frac{\gamma-\epsilon}{1+\epsilon}\alpha
+(1-\gamma)\beta\alpha\leq {\frac { \left( 2\,\sqrt {3}-3 \right)
    \left( e-1 \right) }{3\,e+3 +2\,\sqrt {3}}}\alpha^2+ 2\,{\frac
  {3+\sqrt {3}}{3\,e+3+2\,\sqrt {3}}} \beta^2$$
\end{lemma}

\begin{proof}
Let $f$ be the expression that we take if we substitute $\gamma={\frac
  { \left( 3+2\,\sqrt {3} \right) \left( e-3+2\,\sqrt {3} \right)
  }{3\,e+3+2\,\sqrt {3}}}$, and then substract the first part from the
second and divide by ${\frac { \left( 2\,\sqrt {3}-3 \right) \left(
    e-1 \right) }{3\,e+3 +2\,\sqrt {3}}}$.  We can study $f$ as a
function of integers $\alpha$ and $\beta$.

$$f(\alpha,\beta)=1/4\, \left( 2\,\sqrt {3}+3-4\,b-2\,b\sqrt {3}+2\,a \right) ^{2}+1/8\, \left( 5+3\,\sqrt {3} \right)  \left( 8\,\beta-3-3\,\sqrt {3}
\right).$$
 
We want to prove that $f(\alpha,\beta)\geq 0$, for every
$\alpha,\beta\in \mathbb{N}$.  One can easily verify that
$$f(\alpha,0)=\left( 3+a+2\,\sqrt {3} \right) a\geq 0,$$
$$f(\alpha,1)=\alpha(\alpha-1)\geq 0,$$ while for $\beta\geq 2$ it gets only positive values.
\end{proof}

We can now prove the most important result of this section.
\begin{theorem}[Atomic-PoS-Upper-Bound]\label{thm:pos-upper-atomic}
For any positive real $\epsilon\leq 1$, the approximate price of
stability of general congestion games with linear latencies is at most
$$\frac{\sqrt{3}+1}{\sqrt{3}+\epsilon}.$$
\end{theorem}
\begin{proof}
Let $A$ be the allocation that minimizes the $\epsilon$ potential
$\Phi^{\epsilon}$, and let $P$ be the optimum allocation. Since $A$ is
a local minimum of $\Phi^\epsilon$, if we sum (\ref{eq:deviation}) for
all players $i$, we get
$$
\sum_{e\in E}n_e(A)\left(a_e n_e(A)+\frac{1}{1+\epsilon}(-a_e\epsilon
+b_e)\right)\leq \sum_{e\in E}n_e(P) \left(a_e
n_e(A)+\frac{1}{1+\epsilon}(a_e+b_e)\right) - \sum_{i\in N}\sum_{e\in
  P_i\cap A_i} a_e.$$ For simplicity let's assume $a_e=1, b_e=0$,
although the results hold in general. We get
\begin{equation}\label{eq:eps-potential-nash}
\sum_{e\in E}\left(n_e^2(A)-\frac{\epsilon}{1+\epsilon}n_e(A)\right)\leq \sum_{e\in E}n_e(P)
\left(n_e(A)+\frac{1}{1+\epsilon}\right) 
\end{equation}

If we multiply (\ref{eq:eps-potential}) with $\gamma$ and
(\ref{eq:eps-potential-nash}) with $(1-\gamma)$, for $\gamma={\frac {
    \left( 3+2\,\sqrt {3} \right) \left( e-3+2\,\sqrt {3} \right)
  }{3\,e+3+2\,\sqrt {3}}}$ and add them, we get
\begin{eqnarray*}
\sum_{e\in E}n_e^2(A) &\leq &\gamma\beta^2 + \frac{1-\gamma\epsilon}{1+\epsilon}\sum_{e\in E}n_e(P)-\frac{\gamma-\epsilon}{1+\epsilon}\sum_{e\in E}n_e(A)
+(1-\gamma)\sum_{e\in E}n_e(P)n_e(A)\\
&\leq &{\frac { \left( 2\,\sqrt {3}-3 \right)  \left( 1-\epsilon \right) }{3\,\epsilon+3+2\,\sqrt {3}}}\sum_{e\in E}n_e^2(A)+{\frac {6+2\sqrt {3}}{3\,\epsilon+3+2\,\sqrt {3}}}\sum_{e\in E}n_e^2(P)\\	
\end{eqnarray*}
and so
$$\sum_{e\in E}n_e^2(A) \leq \frac{\sqrt{3}+1}{\sqrt{3}+\epsilon}\sum_{e\in E}n_e^2(P).$$
\end{proof}

Theorem~\ref{thm:potential-atomic} implies that the socially optimal
allocation is $1-$equilibrium. So for $\epsilon\geq 1$, trivially the
price of stability is 1. The following theorem shows that this is
tight, in the sense that the social cost of the best
$(1-\delta)$-approximate equilibrium, is strictly greater than the
social optimum.

\begin{theorem}
There exist instances of congestion games, (even with two parallel
links), where a the optimum allocation is not a
$(1-\delta)-$approximate equilibrium, for any arbitrarily small
positive $\delta$.  This means that the price of stability for
$(1-\delta)$-approximate equilibria is strictly greater than 1.
\end{theorem}

\begin{proof}
Consider a game with two facilities {(\em parallel links)} $e_1, e_2$
and $n$ players.  The facilities have latencies
$l_{e_1}(x)=(2n-1)\cdot x-\gamma$, for some arbitrary small positive
$\gamma$ and $l_{e_2}(x)=x$.

Consider the allocation $P$, that is produced when one player, (say
the first), chooses the first link and the rest of the players use the
second link. This has cost
$$\scsum(P)=2n-1-\gamma+(n-1)^2,$$ which is optimal:
Any other allocation, in which $k\neq 1$ players use the first link
and $n-k$ the second, has cost
$$(2n-1-\gamma)k^2+(n-k)^2\geq(2n-1-\gamma)+(n-1)^2.$$
In strategy profile $P$, the first player has cost $2n-1-\gamma$,
while the rest of the players have cost $n-1$ each.  If the first
player unilaterally deviates to the second link he will have cost $n$.
This means that $\opt$ is a $(1-\frac{1+\gamma}{n})$-approximate
equilibrium. Therefore, for any $\delta$, there is an instance with
sufficiently large number of players $n(\delta)$, where $\opt$ is not
a $(1-\delta)$-approximate equilibrium.
\end{proof}

We now give an almost matching lower bound for the price of
stability. The upper and lower bounds are not equal but they match at
the extreme values of $\epsilon=0$ and $\epsilon=1$.  For
$\epsilon=0$, we get the known price of stability
\cite{CK05b,CFKKM06}. The price of stability decreases as a function
of $\epsilon$, and drops to 1 for $\epsilon = 1$.

\begin{theorem}[Atomic-PoS-Lower-Bound]\label{thm:ps-dominant-lower} 
There are linear congestion games whose approximate Nash equilibria
(even their dominant equilibria as the proof reveals) have price of
stability of the \scsum\ social cost approaching
$$2\,{\frac {3+\epsilon+{\theta}{\epsilon}^{2}+3\,{\epsilon}^{3}+2\, 
{\epsilon}^{4}+{\theta}+{\theta}\epsilon}
  {6+2\,\epsilon+5\,{\theta}\epsilon+6\,{\epsilon
    }^{3}+4\,{\epsilon}^{4}-{\theta}{\epsilon}^{3}+2\,{\theta}{\epsilon}^{2}}},$$
where $\theta=\sqrt {3\,{\epsilon}^{3}+3+\epsilon+2\,{\epsilon}^{4}}$.
\end{theorem}

\begin{proof}
We describe a game of $n+\lambda$ players with parameters $\alpha$,
$\beta$, and $\lambda$ which we will fix later to obtain the desired
properties. Each player $i$ has two strategies $A_i$ and $P_i$, where
the strategy profile $(A_1,\ldots,A_n)$ will be the equilibrium and
$(P_1,\ldots,P_n)$ will have optimal social cost.

There are also $\lambda$ players that have fixed strategies; they
don't have any alternative.  They play a fixed facility $f_{\lambda}$.

There are 3 types of facilities:
\begin{itemize}
\item $n$ facilities $\alpha_i$, $i=1,\ldots,n$, each with cost
function $l(x)=\alpha x$. Facility $\alpha_i$ belongs only to strategy
$P_i$.
\item $n(n-1)$ facilities $\beta_{ij}$, $i,j=1,\ldots,n$ and $i\neq
j$, each with cost $l(x)=\beta x$. Facility $\beta_{ij}$ belongs only
to strategies $A_i$ and $P_j$.
\item 1 facility $f_\lambda$ with unit cost $l(x)=x$.
\end{itemize}
We will first compute the cost of every player and every strategy
profile.  By symmetry, we need only to consider the cost $cost_A(k)$
of player 1 and the cost $cost_P(k)$ of player $N$ of the strategy
profile $(A_1,\ldots,A_k,P_{k+1},\ldots,P_n)$. Therefore,
\begin{eqnarray*}
cost_A(k) &=&(2n-k-1)\beta + (\lambda + k).
\end{eqnarray*}

Similarly, we compute
\begin{eqnarray*}
cost_P(k) &=& \alpha+(n+k-1)\beta.
\end{eqnarray*}
  
We now want to select the parameters $\alpha$ and $\beta$ so that the
strategy profile $(A_1,\ldots,A_n)$ is
$(1+\epsilon)$-dominant. Equivalently, at every strategy profile
$(A_1,\ldots,A_k,$ $P_{k+1},\ldots,P_n)$, player $i$, $i=1,\ldots,k$,
has no reason to switch to strategy $P_i$, because it's $(1+\epsilon)$
times less profitable. We use dominant strategies because it is easier
to guarantee that there is no other equilibrium. This is expressed by
the constraints
\begin{equation}
\label{eq:dominant}
(1+\epsilon)\cdot cost_A(k) \leq
cost_P(k-1),\quad\textrm{for every $k=1,\ldots,n$.}
\end{equation}

All these constraints are linear in $k$ and they are satisfied by
equality when

$$\alpha={\frac { \left( 1+\epsilon \right)  \left( 2\,n\epsilon-\epsilon+\epsilon\lambda+n+2\,\lambda+1 \right) }{2+\epsilon}
}$$ and $$\beta = {\frac {1+\epsilon}{2+\epsilon}},$$
as one can verify with straightforward, albeit tedious, substitution.

In summary, for the above values of the parameters $\alpha$ and
$\beta$, we obtain the desired property that the strategy profile
$(A_1,\ldots,A_n)$ is a $(1+\epsilon)$-dominant strategy. If we
increase $\alpha$ by any small positive $\delta$, inequality
(\ref{eq:dominant}) becomes strict and the $(1+\epsilon)$-dominant
strategy is unique (and therefore unique Nash equilibrium).

We now want to select the value of the parameter $m$ so that the price
of anarchy\footnote{Since this is the unique $1+\epsilon$ Nash
  Equilibrium of this game, the terms price of anarchy and price of
  stability are equivalent.}  of this equilibrium is as high as
possible. The price of anarchy
is $$\frac{cost_A(N)+\lambda(\lambda+n)}{cost_P(0)+\lambda^2}$$ which
for the above values of $\alpha$ and $\beta$ can be simplified to
$${\frac {3\,{n}^{2}+2\,{n}^{2}\epsilon-n-n\epsilon+4\,n\lambda+2\,n\lambda\epsilon+2\,{\lambda}^{2}+\epsilon{\lambda}^{2}}{
    4\,{n}^{2}\epsilon-n\epsilon+3\,n\lambda\epsilon+2\,{n}^{2}+2\,n\lambda+2\,{n}^{2}{\epsilon}^{2}-n{\epsilon}^{2}+n{\epsilon}
    ^{2}\lambda+2\,{\lambda}^{2}+\epsilon{\lambda}^{2}}}.$$ If we
optimize the parameter $\lambda$ and take the limit of $n$ to infinity
we get the theorem.
\end{proof}

\section{Selfish Routing -- PoS}

Here we follow the ideas of the previous section to define an
appropriate potential function for $\epsilon$-Nash equilibrium for the selfish
routing problem or more generally non-atomic congestion games. It is
easier to deal with the more general case of non-atomic congestion
games rather than the selfish routing case, since we don't have to
concern ourselves with the underlying network. In fact, our approach
reveals how little we really need to establish results that encompass
many influential results in the literature.

Consider a flow $f$ for the selfish routing with flow $f_e$ through
every edge $e$. We define the $\epsilon$-potential function
$$\Phi^{\epsilon}(f)=\sum_{e\in E} \left(\frac{1}{2}\, a_e f_e^2 +
\frac{1}{1+\epsilon}\, b_e f_e\right).$$
We will show that the global minimum of $\Phi^{\epsilon}(f)$ is an
$\epsilon$-Nash equilibrium:
\begin{theorem}\label{thm:potential-selfish}
In a non-atomic congestion game, the flow $f$ which minimizes the
$\epsilon$-potential function is an $\epsilon$-Nash
equilibrium. Furthermore, for any other flow $f'$ the following
inequality holds:
\begin{equation*}
\sum_{e\in E}\left(a_e f_e^2+\frac{1}{1+\epsilon}\, b_e f_e\right) \leq
\sum_{e\in E}\left(a_e f_e f_e'+\frac{1}{1+\epsilon}\, b_e f_e'\right).
\end{equation*}
\end{theorem}

\begin{proof}
Consider a flow $f$ and two paths $p$ and $p'$ of the same
commodity. Suppose that the flow $f$ on path $p$ is positive. We want
to compute how much $\Phi^{\epsilon}(f)$ changes when we shift a small
amount $\delta>0$ of flow from path $p$ to path $p'$.  More precisely,
if $f'$ denotes the new flow, we compute the following limit
\begin{equation}
\label{eq:routing_phi}
\lim_{\delta\rightarrow
  0}\frac{\Phi^{\epsilon}(f')-\Phi^{\epsilon}(f)}{\delta}=\sum_{e\in
  p'}\left(a_e f_e+\frac{1}{1+\epsilon}\, b_e\right) - \sum_{e\in
  p}\left(a_e f_e+\frac{1}{1+\epsilon}\, b_e\right)
\end{equation}
If $f$ minimizes $\Phi^{\epsilon}$, then the above quantity is
nonnegative. But we can bound the cost of paths $p$ and $p'$ with the
two terms of this quantity as follows:
$$l_p(f)=\sum_{e\in p} (a_e f_e + b_e) \leq (1+\epsilon) \sum_{e\in p}
(a_e f_e + \frac{1}{1+\epsilon}\, b_e)$$
and
$$l_{p'}(f)=\sum_{e\in p'} (a_e f_e + b_e) \geq \sum_{e\in p'}
(a_e f_e + \frac{1}{1+\epsilon}\, b_e).$$
It follows that $l_p(f)\leq (1+\epsilon) l_{p'}(f)$, which implies
that $f$ is an $\epsilon$-Nash equilibrium.

For the second part, we observe that the expression
(\ref{eq:routing_phi}), which is nonnegative for $f$ which minimizes
$\Phi^{\epsilon}$, implies that for every path $p$ on which $f$ is
positive and every other path $p'$ we must have
$$ \sum_{e\in p}\left(a_e f_e+\frac{1}{1+\epsilon}\, b_e\right) \leq \sum_{e\in
  p'}\left(a_e f_e+\frac{1}{1+\epsilon}\, b_e\right).$$ Consider now
another flow $f'$ which satisfies the rate constraints for the
commodities and let us sum the above inequalities for all paths $p$
and $p'$ weighted with the amount of flow in $f$ and $f'$. More
precisely:
\begin{align*}
\sum_{p,p'} f_p f_{p'}' \sum_{e\in p} \left(a_e
f_e+\frac{1}{1+\epsilon}\, b_e\right) \leq \sum_{p,p'} f_p f_{p'}'
\sum_{e\in p'} \left(a_e
f_e+\frac{1}{1+\epsilon}\, b_e\right) \\
\sum_{p'} f_{p'}' \sum_{e\in E} \left(a_e f_e^2+\frac{1}{1+\epsilon}\,
b_e f_e\right) \leq \sum_{p} f_p \sum_{e\in E} \left(a_e f_e f_e' +
\frac{1}{1+\epsilon}\, b_e f_e'\right)
\end{align*}
But $\sum_{p'} f_{p'}'=\sum_p f_p$ is equal to the sum of the rates
for all commodities. If we remove from the expression this common
factor, the second part of the theorem follows.
\end{proof}

From Lemma~\ref{lem:poa-upper-selfish}, if we substitute $\lambda$ with
$1/(1+\epsilon)$, we get that for any reals $\alpha, \beta$, and
$\epsilon \in [0,1]$

\begin{equation}\label{eq:real-inequality-pos}
\alpha\beta\leq \frac{1+\epsilon}{4}\alpha^2 + \frac{1}{1+\epsilon}\beta^2.
\end{equation}

\begin{theorem}[Selfish-PoS-Upper-Bound]
The price of stability is at most $\frac{4}{(3-\epsilon)(1+\epsilon)}$.
\end{theorem}
\begin{proof}
Let $f$ be the potential minimizer of $\Phi^\epsilon$ and $f^*$ be the
optimum flow. From Theorem (\ref{thm:potential-selfish}) and
(\ref{eq:real-inequality-pos}) we get that

$$\sum_{e\in E}a_ef_e^2+\frac{1}{1+\epsilon}b_ef_e \leq \sum_{e\in E}a_e(\frac{1+\epsilon}{4}{f_e}2 + \frac{1}{1+\epsilon}{f_e^*}^2)+\frac{1}{1+\epsilon}b_ef_e^*$$
or
$$\sum_{e\in E}a_e\frac{3-\epsilon}{4}f_e^2+\frac{1}{1+\epsilon}b_ef_e \leq \frac{1}{1+\epsilon}C(f^*),$$
and since $1/(1+\epsilon)\geq (3-\epsilon)/4$, we get
$$\frac{3-\epsilon}{4}C(f)\leq \frac{1}{1+\epsilon}C(f^*),$$
which gives the desired result:
$$C(f)\leq \frac{4}{(3-\epsilon)(1+\epsilon)}C(f^*).$$
\end{proof}

We now establish that the Pigou network (extended to take into account
the parameter $\epsilon$, Figure~\ref{fig:pigou}) gives tight results.
\begin{theorem}[Selfish-PoS-Lower-Bound]
The price of stability is at least $\frac{4}{(3-\epsilon)(1+\epsilon)}$.
\end{theorem}

\begin{proof}
Consider the Pigou network of Figure~\ref{fig:pigou}. There is a unit
of flow that wants to move from $s$ to $t$. Clearly, the only
$(1+\epsilon)$-Wardrop flow is to choose the lower edge, for
$\epsilon<1$.  This gives a social cost of $1$.

On the other hand the optimum is to route $(1+\epsilon)/2$ of the
traffic from the lower edge and $(1-\epsilon)/2$ of the traffic from
the upper edge.  This gives a social opt of
$\frac{(1+\epsilon)(1-\epsilon)}{2}+\frac{(1+\epsilon)}{2}\frac{(1+\epsilon)}{2}=\frac{(1+\epsilon)(3-\epsilon)}{4}$,
and so the price of stability is $\frac{4}{(3-\epsilon)(1+\epsilon)}$
as needed.
\end{proof}

\paragraph{Acknowledgements}
The authors would like to thank Ioannis Caragiannis for many helpful
discussions and Tim Roughgarden for useful pointers to literature.


\bibliographystyle{plain}

\end{document}